\documentclass{article}
\usepackage[utf8]{inputenc}
\usepackage{authblk}
\usepackage{indentfirst}
\usepackage{hyperref}
\usepackage{longtable}
\usepackage{setspace}
\usepackage[margin=1in]{geometry}
\usepackage{graphicx}
\graphicspath{ {./figures/} }
\usepackage{subcaption}
\usepackage{amsmath}
\usepackage{lineno}
\usepackage{multirow}
\usepackage{color,soul}
\usepackage[version=3]{mhchem}

\usepackage[biblabel]{cite}

\title{Probing Structure and Ionic Transport in Molten Lithium Carbonate}

\author[1]{Debsundar Dey}
\author[2]{Abhirup Patra}
\author[3]{Anand Narayanan Krishnamoorthy}
\author[1,*]{Gopalakrishnan Sai Gautam}

\affil[1]{Department of Materials Engineering, Indian Institute of Science, Bengaluru, 560012, India}
\affil[2]{Shell International Exploration \& Production Inc., 200 N Dairy Ashford Rd, Houston, TX 77079, United States}
\affil[3]{Shell India Markets Pvt. Ltd. (Shell Projects \& Technology), Mahadeva Kodigehalli, Bengaluru 562149, Karnataka, India}
\affil[*]{Email: \href{mailto:abhirup.patra@shell.com}{abhirup.patra@shell.com}; \href{mailto:saigautamg@iisc.ac.in}{saigautamg@iisc.ac.in}}

\date{}

\begin{document}

\maketitle

%%%%%% Abstract %%%%%%
\begin{abstract}
\noindent Lithium carbonate (Li$_2$CO$_3$) is a cornerstone material for clean energy technologies, including high-temperature molten carbonate fuel cells, electrochemical carbon capture, and lithium-based batteries. However, capturing the complex, many-body interactions governing the structure and transport in Li$_2$CO$_3$, especially in its molten state, has remained a challenge, constrained by the computational cost of \textit{ab initio} methods and the accuracy limitations of classical force fields. To address this gap, we deploy equivariant graph-based machine learned interatomic potentials, specifically, the multi atomic cluster expansion (MACE) and neural equivariant interatomic potential (NequIP) architectures that are trained on melt-quench \textit{ab initio} molecular dynamics data. Our benchmarking demonstrates that MACE, utilizing higher body-order message passing, provides superior transferability and precision in predicting energies and forces compared to NequIP. Subsequently, we use the optimized MACE model to perform large-scale molecular dynamics simulations to probe the structural and transport properties of molten $\text{Li}_2\text{CO}_3$. Besides describing the structural features, such as the dominant presence of C-O pair correlations under molten conditions, our MACE model also reproduces experimentally-measured static structure factors and shear viscosity values. Further, our simulations indicate that Li$^+$ transport in Li$_2$CO$_3$ is fundamentally dominated by concerted motion, as evidenced by Haven's ratios being significantly below unity ($\approx 0.20-0.40$). Notably, we identify a temperature-driven transition from anisotropic (and highly concerted) Li$^+$ transport, supported by persistent oxygen-centered Voronoi cages at 1000~K, to isotropic (and less concerted) diffusion at 1400 K. Thus, we provide fundamental insights into the structural and transport properties of molten Li$_2$CO$_3$ and also demonstrate a robust and scalable framework for the accelerated design of molten salt electrolytes and ionic liquids for clean energy applications.
\end{abstract}

%%%%%% Main Text %%%%%%

\subsection*{Introduction}
\noindent Molten alkali carbonates represent a critically important class of ionic liquids with wide-ranging technological applications in clean energy systems \cite{intro1}. For example, the high-temperature ionic melts serve as electrolytes in molten carbonate fuel cells (MCFCs), which operate at temperatures of 600- 700~°C and achieve electrical efficiencies approaching 60\%, with combined heat and power efficiencies reaching 85\% \cite{McPhail2008_MCFC_Status,GhezelAyagh2017_MCFC_Advances}. Among the alkali carbonates, lithium carbonate (Li$_2$CO$_3$) plays a particularly prominent role due to its superior ionic conductivity compared to sodium and potassium carbonates \cite{Kojima2008_TernaryCarbonateProps,Janz1988_MoltenSaltData}. Standard MCFC electrolytes typically contain 62$-$68~mol\% Li$_2$CO$_3$ mixed with K$_2$CO$_3$ or Na$_2$CO$_3$, with the lithium component providing enhanced ionic transport while the secondary carbonate reduces corrosion and improves gas solubility \cite{Barckholtz2021_CO3_OH_Equilibrium,Frangini2013_MCFC_Chapter}. Beyond fuel cells, molten Li$_2$CO$_3$ has emerged as a promising medium for electrochemical CO$_2$ capture and subsequent conversion to valuable carbon nanomaterials, including carbon nanotubes and graphene nanocarbons, offering a pathway for large-scale decarbonization \cite{Licht2010_STEP_CarbonCapture,Ren2024_NewElectrolyte_MC_Decarbonization}.

Despite their technological importance, achieving a fundamental, atomistic understanding of the structure and dynamics of molten carbonates remains challenging. The extreme chemical reactivity of these melts at elevated temperatures, combined with their corrosive nature, presents significant obstacles for experimental characterization of atomic structures and transport properties \cite{Bini2007_IL_InteractionScales,Fiorito2020_MoltenCarbonates_Transport}. Experimental techniques such as X-ray and neutron scattering can provide structural information through the static structure factor ($S(q)$) and radial distribution functions (RDFs, denoted by $G(r)$ or $g(r)$), while electrochemical methods and pulsed-field gradient nuclear magnetic resonance can probe transport coefficients \cite{Wilding2016_LowDimNetwork_Na2CO3,Sessa2024_LiKCO3_Polarization,Takeuchi2007_Structure_MoltenLi2CO3}. However, these measurements are technically demanding and provide limited insight into the microscopic mechanisms governing ionic transport, particularly the nature of correlated ionic motion that deviates from the ideal Nernst-Einstein behavior. Resolving whether ionic transport in these melts follows a traditional, independent random-walk model or a more complex, many-body correlated mechanism, and the local structural motifs that influence macroscopic conductivity are essential for the optimal design of high-performance electrolytes that enhance ionic conductivity and reduce interfacial resistance.

Beyond their importance in molten carbonate systems, lithium carbonates are also technologically relevant in Li-ion batteries\cite{SEI7,jagger2023solid,adenusi2023lithium,wang2019identifying}. Li$_2$CO$_3$ is one of the dominant inorganic constituents of the solid electrolyte interphase (SEI), where it forms crystalline and amorphous domains that govern Li$^+$ transport across the electrode-electrolyte interface\cite{SEI1,SEI2,SEI3,peled2017sei}. The SEI stabilizes the anode by preventing continuous electrolyte decomposition, but it also introduces additional interfacial resistance, making Li$^+$ mobility through Li$_2$CO$_3$-rich domains a critical determinant of fast-charging performance and long-term cycling stability\cite{SEI4,SEI5,SEI6}. The contrasting Li$^+$ conductivities of crystalline and amorphous Li$_2$CO$_3$, as well as the mixed-phase and polydomain character of realistic SEIs, highlight the broader importance of understanding Li ionic transport in Li$_2$CO$_3$ across both solid and molten regimes.

Atomistic simulation provides a powerful complementary approach for elucidating the structure-property relationships in molten salts. Classical molecular dynamics (MD) simulations based on empirical force fields, such as rigid-ion or polarizable shell models, have been widely employed to study molten carbonates \cite{Tosi1964_IonicSizes_BornParams,Mondal2020_GA_ForceField_Carbonates,Mills1973_SelfDiffusion_Water}. While computationally efficient, these classical potentials often struggle to accurately capture the complex many-body interactions and polarization effects that are critical in ionic systems. \textit{Ab initio} molecular dynamics (AIMD) based on density functional theory (DFT) offers quantum mechanical accuracy but is computationally prohibitive for accessing the long time scales (hundreds of picoseconds to nanoseconds) and large system sizes (thousands of atoms) required for converged transport properties and proper sampling of the liquid structure \cite{CarParrinello1985_CP_MD,MarxHutter2009_AIMD_Book}. This fundamental trade-off between accuracy and computational efficiency has motivated the development of machine learned interatomic potentials (MLIPs) that can achieve near-DFT accuracy at a fraction of the computational cost of AIMD simulations.

Various MLIP architectures have been successfully applied to molten salts and related systems, including neural network potentials, Gaussian approximation potentials (GAP), moment tensor potentials (MTP), and deep potential molecular dynamics (DeepMD)  \cite{BehlerParrinello2007_NN_PES,Tovey2020_MoltenNaCl_ML,Bartok2010_GAP_PRL,Sivaraman2021_MoltenLiCl_ML,Shapeev2016_MTP_MMS,Zhang2018_DeepPotential_PRL}. Similar successes have been reported for diverse molten salt chemistries relevant to energy conversion, nuclear applications, and high-temperature electrochemical technologies \cite{Sivaraman2021_MoltenLiCl_ML,Lam2021_LiF_FLiBe_NNIP,Nguyen2021_ActinideMoltenSalts}. Recently, MLIPs have also been used for characterizing ionic transport across amorphous materials and their interfaces in battery systems \cite{choyal2025exploration,seth2025investigating}. So far, equivariant MLIPs have not been used to probe the transport and local structure motifs in molten Li$_2$CO$_3$.

Recent advances in equivariant neural network architectures have further improved the accuracy and data efficiency of MLIPs. For example, E(3)-equivariant graph-based neural networks, which explicitly include the symmetries of three-dimensional space (i.e., translations, rotations, and reflections), along with message passing, provide a better representation of atomic environments compared to invariant neural networks \cite{Geiger2022_e3nn,Batzner2023_Review_EquivariantIPs}. Specifically, the neural equivariant interatomic potential (NequIP) architecture employs E(3)-equivariant convolutions to describe geometric tensors, achieving high accuracy with remarkable data efficiency that outperforms invariant approaches with up to three orders of magnitude fewer training data \cite{Batzner2022_NequIP_NatCommun}. Building upon NequIP, the multi atomic cluster expansion (MACE) architecture introduces higher body-order message passing, enabling the capture of many-body interactions (typically up to four-body terms). By explicitly accounting for such many-body interactions, MACE is uniquely positioned to resolve the subtle, concerted ion-ion interactions that govern transport in molten carbonates—features that are often missed by traditional invariant potentials \cite{Batatia2022_MACE_NeurIPS,batatia2025design}. Given the utility in incorporating equivariance, message passing and many-body interactions (in MACE), both NequIP and MACE have been successfully applied to challenging systems, such as, ionic liquids, water, amorphous solids, and complex molecular systems \cite{Goodwin2024_IonicLiquids_EquivariantMLIP,Batatia2023_MACE_JCP,Musaelian2023_LocalEquivariantReps,seth2025investigating, tenti2025hydrogen,devereux2024performance,beiersdorfer2025gap}.

In this work, we construct and assess equivariant MLIPs and subsequently probe the structural and transport properties in molten Li$_2$CO$_3$. We construct the MLIPs based on AIMD training data, with training configurations sampling melt–quench trajectories at 1500~K and 1250~K, resulting in 3600 and 1988 structures, respectively. To assess model fidelity against AIMD and available experimental data, we use RDFs and $S(q)$ as metrics. We train both the MACE and NequIP architectures on our dataset, from scratch, and identify MACE to be the best model based on the train and test mean absolute errors (MAEs) across energies and forces. Additionally, we evaluate the medium-sized MACE-MP-0 model \cite{batatia2025foundation}, a foundational pre-trained MLIP without fine-tuning, to benchmark structural accuracy. Using the best-performing potential, i.e., MACE architecture trained from scratch, simply referred to as MACE in the remainder on the text, we perform long time-scale MD simulations to compute macroscopic transport properties, including the shear viscosity and Li$^+$ self-diffusion coefficients. 

Importantly, we find the activation energy of Li$^+$ diffusion ($E_a$), calculated based on an Arrhenius fit of Li$^+$ diffusivities  over temperatures to be $1.287$~eV, which is consistent with prior molten-salt literature \cite{SEI6}. Our analysis of ionic trajectories reveals distinct temperature-dependent diffusion mechanisms. For example, at 1400~K, Li$^+$ exhibit continuous migration through the molten carbonate network, while at 1000~K transient trapping events emerge, wherein Li$^+$ oscillate between nearby carbonate coordination cages before escaping. Large-scale simulations enabled by the optimized MACE potential further show that Li-ion transport in molten Li$_2$CO$_3$ is fundamentally collective, with Haven's ratio ($H_R$) values significantly below unity ($H_R \sim$ 0.20–0.40). Finally, we identify a temperature-driven transition from anisotropic transport at 1000~K to isotropic diffusion at 1400~K, demonstrating the ability of equivariant MLIPs to resolve complex transport behavior across different molten carbonate regimes. We hope that our work is a useful contribution in the optimal design of molten salt electrolytes for MCFCs and related applications.

%\newpage
%\section{Methods}
\subsection*{Dataset Generation and Optimized MLIPs}
\begin{figure}[h!]
\centering
\includegraphics[width=1\textwidth]{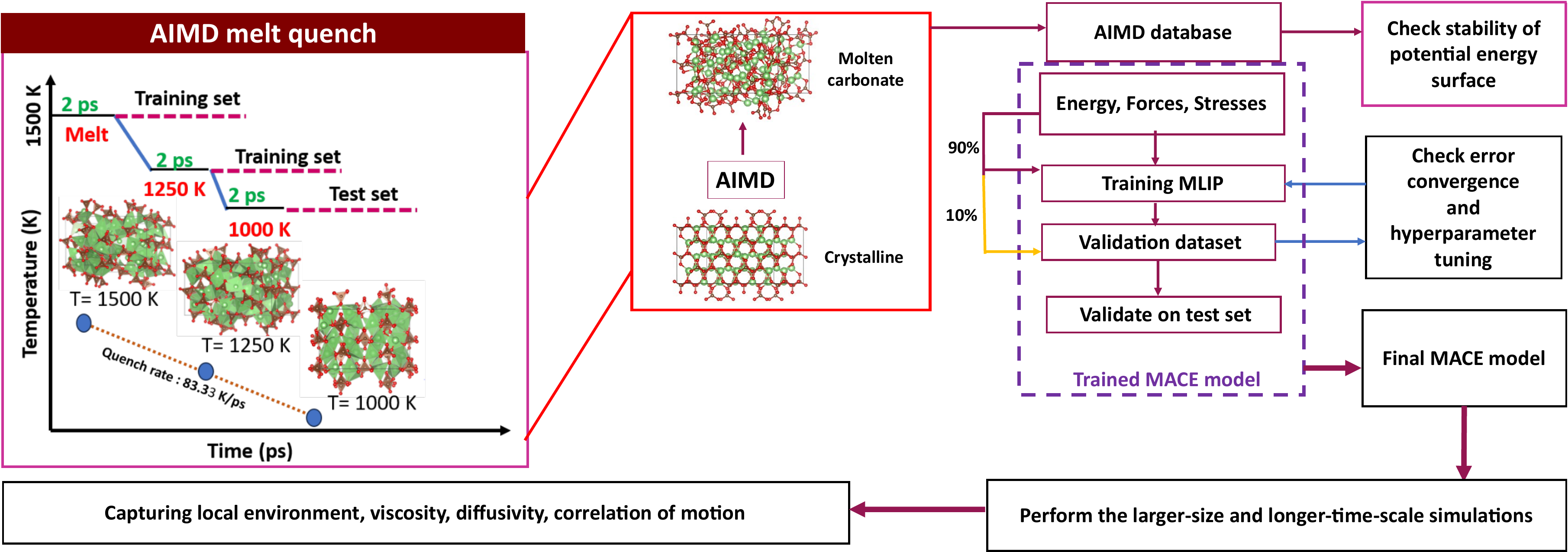}
\caption{Workflow for dataset generation and MLIP training (for MACE) employed in this work. We employ AIMD melt-quench simulations both for training and hyperparameter optimization. The accuracy and transferability of the trained MLIP are evaluated using independent test configurations. After validation, the optimized MACE model is deployed for long time-scale and large system-size MD simulations to characterize transport properties of molten-Li$_2$CO$_3$.}
\label{fig:workflow}
\end{figure}

\noindent \textbf{Figure~{\ref{fig:workflow}}} summarizes the overall workflow of this study, including AIMD-based data generation, MLIP training and validation, and subsequent molten-phase simulations performed using the optimized model. To develop and evaluate MLIPs for molten-Li$_2$CO$_3$, we generated a dataset using AIMD melt-quench simulations \cite{nayak2025accurate,scopel2008amorphous} on a $2\times2\times2$ supercell (192 atoms) of the crystalline Li$_2$CO$_3$ structure that was obtained from the inorganic crystal structure database (ICSD\cite{icsd}). We applied stepwise heating initially from 0~K to 500~K, subsequently from 500~K to 1000~K, and finally from 1000~K to 1500~K, with each heating step conducted over 5~ps followed by 2~ps of equilibration. The choice of 1500~K as the highest temperature in our heating simulations is motivated by the fact that Li$_2$CO$_3$ melts at 996~K \cite{yan2020experimental}. Subsequently, we quenched the equilibrated structure at 1500~K to 1000~K, in steps of 250~K at a quench rate of 83.33 K/ps. For training, we sampled configurations from AIMD trajectories at 1500~K (accounting for 3600 structures sampled over 7.2 ps) and at 1250~K (1988 structures sampled over 3.98 ps), with the training data collected after the initial 2~ps of equilibration at each temperature. 

We trained both NequIP and MACE models using the dataset generated from the AIMD melt-quench simulations. We performed training and hyperparameter optimization on the AIMD-derived training dataset, by creating a 90:10 train:validation split of the dataset. The list of optimal hyperparameters that were chosen to minimize errors on the validation set for both models are listed in \textbf{Tables~S1} and \textbf{S2} of the supporting information (SI). To ensure a fair comparison, we set the equivariance level of both models to one (i.e., $L_{\mathrm{max}} = 1$). We evaluated the performance of both models using MAEs across both total energies and atomic forces. We used two separate test datasets to evaluate the transferability of the trained MLIPs: (i) configurations from AIMD, post quench, at 1000~K, and (ii) mechanically strained structures of crystalline-Li$_2$CO$_3$, including homogeneous volumetric strain ($\pm 2\%$, $\pm 5\%$, $\pm 10\%$), uniaxial strain ($\pm 2\%$, $\pm 5\%$, $\pm 10\%$), and shear strain (2\%, 5\%, 10\%). To verify that the sampled configurations correspond to the molten phase of Li$_2$CO$_3$, we computed the RDFs for all relevant atomic pairs ($G(r)$), namely O-O, Li-Li, C-O, Li-O, C-C, and Li-C.

\subsection*{DFT and MD calculations }
\noindent We utilized the Vienna \textit{ab initio} simulation package \cite{kresse1996efficiency,vasp2} with projector augmented wave potentials \cite{kresse1999ultrasoft} to perform all spin-polarized DFT and AIMD calculations. We used the Perdew-Burke-Ernzerhof functionalization of the generalized gradient approximation \cite{perdew1996generalized} to account for the electronic exchange and correlation. We applied a kinetic energy cutoff of 520~eV, and sampled the irreducible Brillouin zone using $\Gamma$-centered Monkhorst-Pack $k$-point meshes with a density of at least $32$/\AA{} \cite{monkhorst1976special}. We relaxed the as-obtained crystalline Li$_2$CO$_3$ structure (from the ICSD) by optimizing the cell volume, cell shape, and atomic positions without imposing symmetry constraints before performing AIMD simulations and generating the strained structure dataset. We set the convergence criteria for total energy and atomic forces to $10^{-5}$~eV and $|0.03|$~eV/\AA{}, respectively. For generating the strained structure dataset, we deformed the DFT-relaxed Li$_2$CO$_3$ lattice with the corresponding amount of strain and evaluated the total energy using a single self-consistent field calculation (with energy convergence set to $10^{-5}$~eV) without any further relaxation of the structure.

We performed AIMD simulations on a $2\times2\times2$ supercell of Li$_2$CO$_3$, corresponding to 192 atoms, using a 2~fs time step to ensure accuracy and efficiency. We did classical MD simulations with our trained MACE model using the large scale atomic/molecular massively parallel simulator (LAMMPS) \cite{thompson2022lammps}. We employed a 2~fs time step for the classical MD simulations and used a $4\times4\times4$ supercell containing 1536 atoms. All AIMD and MD runs employed the NVT ensemble with a Nos\'e-Hoover thermostat \cite{hoover1985canonical}, the `\texttt{SMASS}' tag set to 0, and the velocity-Verlet integration scheme \cite{verlet1967computer}. For MD simulations with the MACE model trained from scratch, we equilibrated the structures in the NVT ensemble for 10~ps, followed by 700~ps of production runs to estimate diffusivities at 1500~K, 1400~K, 1300~K, 1250~K and 1000~K. The variations in potential energy (in eV), as estimated by MACE during the 10~ps equilibration period, at 1000~K, 1250~K, 1300~K, and 1400~K are shown in \textbf{Figure~S3}.

\subsection*{Estimation of diffusivity, ionic correlation, and viscosity}
\noindent In molten Li$_2$CO$_3$, ionic conductivity ($\sigma$) and diffusivity ($D(c)$) of any species, such as Li$^+$, are connected through the Nernst-Einstein (NE) relation as,
\begin{equation}
\sigma = \frac{q^2\, c D(c)}{k_{\mathrm{B}} T}
\end{equation}
where $q$, $c$, $k_{\mathrm{B}}$, and $T$ represent the ion’s charge (+1 in case of Li$^+$), the ion's concentration, the Boltzmann constant, and temperature, respectively. While the NE relation provides an estimate of $\sigma$ assuming uncorrelated ionic motion, molten salts generally exhibit strong many-body interactions, leading to significant correlations. Therefore, the calculated $\sigma$ from NE can systematically deviate from the true $\sigma$, with the correlation effects requiring further quantitative analysis, such as using $H_R$ \cite{marcolongo2017ionic}. $D(c)$ in the NE framework describes the ionic flux according to Fick’s first law \cite{fick1855v} and can be expressed as $D_c = \Theta D_J$, where $\Theta$ is the thermodynamic factor. The jump diffusivity ($D_J$\cite{he2018statistical,gautam2019theoretical}) measures the mean squared displacement of the center-of-mass of the mobile ions (denoted as 'MSCD' in our work), thus capturing cross-correlations among individual ionic hops over time. Mathematically, $D_J$ is defined as,
\begin{equation}
D_J = \lim_{t \to \infty} \left[ \frac{1}{2 d t} \frac{1}{N} \left( \sum_{i=1}^{N} \mathbf{r}_i(t) \right)^2 \right]
\end{equation}
where $d$, $t$, and $N$ represent the dimensionality of the diffusion process, time, and number of mobile ions, respectively. $r_i (t)$ indicates the displacement of the $i^{\mathrm{th}}$ ion over $t$.

On the other hand, the tracer diffusivity ($D^*$) tracks the motion of individual ions thereby not accounting explicitly for any cross-correlations among their motion. Specifically, $D^*$ quantifies the mean squared displacement (MSD) that is averaged over $N$ and $t$ and is mathematically defined as, 
\begin{equation}
D^* =  \lim_{t \to \infty} \left[  \frac{1}{2 dt} \frac{1}{N}
\sum_{i=1}^{N}
{[\mathbf{r}_i(t)]^2} \right]
\end{equation}
Note that $D_J$ and $D^*$ can be obtained from the slopes of the linear regimes of MSCD and MSD values, respectively, with respect to $t$ \cite{he2018statistical}. Finally, $H_R$ is defined as the ratio of $D^*$ and $D_J$ as,
\begin{equation}
H_R = \frac{D^*}{D_J}
\end{equation}
Thus, $H_R$ provides insight into how cross-correlations alter the overall ionic mobility. If a system exhibits completely uncorrelated ionic motion, $H_R = 1$. In systems with significant ion–ion interactions, $H_R$ deviates from 1, indicating the presence of cross-correlation. Note that in molten Li$_2$CO$_3$, we do expect cross-correlation effects to be significant since Li$^{+}$ transport does not occur in isolation and is often influenced by the rearrangements in the carbonate network and the local charge neutrality.

Due to the limited total simulation time ($t_{\mathrm{tot}}$) in both AIMD and classical MD, MSD data often exhibits noise, leading to fluctuations in the estimated slope versus $t$. To reduce this statistical variability, we employed the total mean-squared displacement (TMSD, Equation~{\ref{eq:tmsd}}), which averages the MSD over multiple time intervals ($\Delta t$), following the approach of He et al. \cite{he2018statistical}.
\begin{equation}
\mathrm{TMSD}(\Delta t) =
\sum_{i=1}^{N}
\left[
\frac{1}{N_{\Delta t}}
\sum_{t = 0}^{t_{\mathrm{tot}} - \Delta t}
\left|
\mathbf{r}_i(t + \Delta t) - \mathbf{r}_i(t)
\right|^2
\right]
\label{eq:tmsd}
\end{equation}
Here, $N_{\Delta t}$ denotes the number of possible time intervals of duration $\Delta t$. $D^*$ is subsequently determined from the slope of MSD versus $\Delta t$, subject to ranges of $\Delta t$ values for which MSD behaves linearly.
\begin{equation}
D^* = \frac{\mathrm{MSD}(\Delta t)}{2 d \Delta t}
\end{equation}
where $\mathrm{MSD}(\Delta t) = \mathrm{TMSD}(\Delta t)/N$. A similar approach can be used for estimating $D_J$ as well \cite{he2018statistical}. In this work, we set $\Delta t = 50$~ps for all diffusivity calculations because it provides a sufficiently long window to access diffusive behavior while maintaining adequate averaging statistics within the total trajectory length ($t_{\mathrm{tot}} = 700$~ps). Also, the MSD exhibits linear scaling with $\Delta t$ in this regime. We used the \texttt{DiffusionAnalyzer} class of \texttt{pymatgen} \cite{mo2012first,ong2013phase} for post-processing our AIMD and classical MD calculations and determining $D_J$, $D^*$ and $H_R$.

Finally, in an ideal transport regime, where ionic hops occur with uniform frequency and are unaffected by local compositional fluctuations or structural heterogeneity, the overall diffusion coefficient ($D_J$ or $D^*$, denoted as $D$ below) takes the Arrhenius form,
\begin{equation}
D = D_0 \exp\left( -\frac{E_a}{k_{\mathrm{B}} T} \right)
\end{equation}
The pre-exponential factor $D_0$ encapsulates parameters such as the hop distance, the atomic vibrational frequency, the available diffusion-carrier concentration, the correlation factor (which describes deviations from random walk), and the geometric factor (connectivity of pathways). The activation energy ($E_a$) for migration, a material-specific property, governs $D$ exponentially and can be extracted from the slope of $\ln D$ versus $1/T$. Although $D$ may depend on correlation and geometric factor, $E_a$ is typically the dominant rate-limiting parameter and therefore a key quantity for assessing ionic transport performance.

For the calculation of the shear viscosity ($\eta$), we equilibrated the Li$_2$CO$_3$ system in the NVT ensemble for 50~ps, and collected the stress components collected during a 30~ps production run. Note that the viscosity is related to the time integral of the autocorrelation function of the off-diagonal stress-tensor components, within the Green-Kubo formalism\cite{plumari2012shear,desmaele2019atomistic}, as,
\begin{equation}
\eta =
\frac{V}{3k_{\mathrm{B}}T}
\int_{0}^{\infty}
\left[
\left\langle P_{xy}(0)P_{xy}(t) \right\rangle
+
\left\langle P_{xz}(0)P_{xz}(t) \right\rangle
+
\left\langle P_{yz}(0)P_{yz}(t) \right\rangle
\right] dt .
\end{equation}
Here, $V$ is the simulation cell volume, $k_{\mathrm{B}}$ is the Boltzmann constant, $T$ is the temperature. $P_{\alpha\beta}$ represent the off-diagonal components
of the pressure or stress tensor, where $\alpha\beta = xy, xz,$ or $yz$. The angular brackets denote averaging over time.

%\section{Results}
\subsection*{Model comparison}
\noindent To identify the best performing model, among NequIP and MACE, we compared the performance of both models on the training, validation, and test datasets. The parity plots for the MACE (panel \textbf{a}) and NequIP (panel \textbf{b}) models on the test AIMD dataset at 1000~K are shown in \textbf{Figure~S1}, while \textbf{Figure~S2} (panel \textbf{a} for MACE and panel \textbf{b} for NequIP) displays the corresponding results for the strained structures (test) dataset. For NequIP, the training energy MAE is 1.3~meV/atom and the validation energy MAE is 0.3~meV/atom. The corresponding force MAEs are 22.4~meV/\AA{} and 0.31~meV/\AA{} for the training and validation sets, respectively. In contrast, the MACE model exhibits superior accuracy, achieving training and validation energy MAEs of 0.1~meV/atom and force MAEs of 2.4~meV/\AA{} and 3.3~meV/\AA{}, respectively. 

Evaluation on the independent test dataset at 1000~K further highlights the higher accuracy and improved transferability of the MACE model, with energy and force MAEs of 0.2~meV/atom and 0.17~meV/\AA{}, respectively, compared to 2.6~meV/atom and 18.1~meV/\AA{} for the NequIP model. For mechanically strained test structures, the MACE model achieves MAEs of 6.8~meV/atom for energies and 13.16~meV/\AA{} for forces, while NequIP yields higher errors, with energy and force MAEs of 21.08~meV/atom and 24.49~meV/\AA{}, respectively. These results collectively demonstrate the enhanced accuracy and transferability of the MACE model compared to NequIP across the diverse evaluation scenarios considered. Accordingly, we employ the optimized MACE model for all subsequent large-scale and long time-scale MD simulations of molten Li$_2$CO$_3$.

\subsection*{Structural properties of molten Li$_2$CO$_3$}
\begin{figure}[htbp]
\centering
\includegraphics[width=1\textwidth]{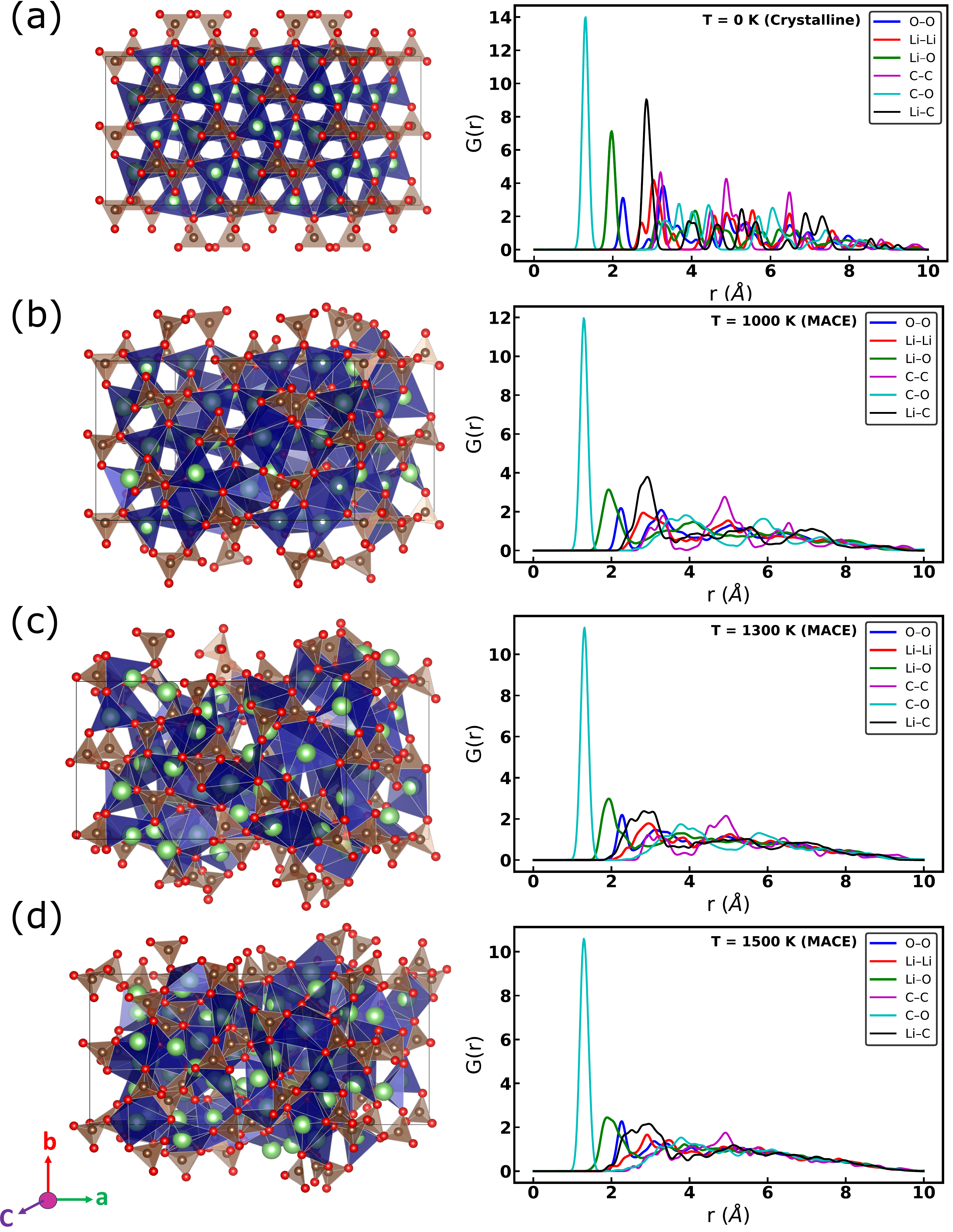}
\caption{Representative atomic configurations (left panels) and corresponding RDFs ($G(r)$; right panels) of Li$_2$CO$_3$ at different temperatures. Panel a corresponds to crystalline Li$_2$CO$_3$ at 0~K relaxed using DFT, while the remaining panels show $G(r)$ for molten configurations generated using the optimized MACE potential at 1000~K, 1300~K, and 1500~K. Blue, red, green, magenta, cyan, and black curves correspond to $G(r)$ of O-O, Li-Li, Li-O, C-C, C-O, and Li-C pairs, respectively.}    
\label{fig:rdfs}
\end{figure}

\noindent We used RDFs and $S(q)$ to characterize the structural properties of molten Li$_2$CO$_3$. Note that $S(q)$ plays a particularly important role, as it is directly accessible via experimental techniques such as X-ray and neutron scattering. We examine the RDFs of all atomic pairs to resolve the short- and medium-range structural ordering within Li$_2$CO$_3$, as compiled in \textbf{Figure~{\ref{fig:rdfs}}}. Note that we compiled RDFs of crystalline Li$_2$CO$_3$ (calculated with DFT), and molten Li$_2$CO$_3$ (above 1000~K, calculated with MACE) in \textbf{Figure~{\ref{fig:rdfs}}}, with some complimentary RDF data generated by AIMD and MACE compiled in \textbf{Figures~S4} and \textbf{S5}. The predicted RDFs by the NequIP and MACE-MP-0 models are provided in \textbf{Figures~S7-S9}).

For crystalline Li$_2$CO$_3$ (\textbf{Figure~{\ref{fig:rdfs}}a}), the DFT-calculated RDFs exhibit sharp and well-defined peaks, with distinct peaks appearing over long distances reflecting the presence of long-range periodicity in the lattice. After melt-quench to 1000~K and subsequent heating to 1500~K (with MACE, panels \textbf{b-d} of \textbf{Figure~{\ref{fig:rdfs}}}), the RDF peaks progressively broaden and decrease in intensity with increasing temperature, indicating a gradual loss of long-range structural order and increased configurational disorder in the molten state. Indeed, the atomic configurations reveal a distinct progression from a periodic network of well-defined Li-O polyhedra in the crystalline phase (\textbf{Figure~{\ref{fig:rdfs}}a}) to a fully disordered molten structure (\textbf{Figure~{\ref{fig:rdfs}}d}), wherein the Li-centered coordination environments become increasingly distorted and non-uniform. We observe an exception to the trend of broadening RDF peaks for the C-O pair (cyan peak), where we find only a marginal reduction in the height of the first coordination peak, even at 1500~K. The lack of significant changes in the C-O peak highlights the strong covalent character of the C-O bond that preserves the planar carbonate geometry irrespective of temperature. Note that although the carbonate units remain locally intact across all temperatures, their orientations become progressively random with increasing temperature, resulting in the loss of medium- and long-range structural order.

At 1000~K (\textbf{Figure~{\ref{fig:rdfs}}b} and \textbf{Figure~S4a}), both MACE and AIMD calculations show pronounced ordering in Li$_2$CO$_3$, with a strong C-O peak at $\sim$1~\AA{}, followed by additional peaks at $\sim$4~\AA{} and $\sim$6~\AA{}, although these higher-order peaks are significantly reduced in intensity compared to the crystalline state. Beyond $\sim$6~\AA{}, no sharp peaks are observed, consistent with the loss of long-range periodicity. The Li-O pairs (green) also show significant peak broadening beyond $\sim$2~\AA{}, while Li-C pairs (black) retain some medium-range ordering. Further, we compare the C-O RDF predicted by the DeepMD model of Dina \textit{et al.} \cite{kussainova2023molecular} at 1043.15~K with our MACE data at 1000~K, and find that we accurately reproduce the peak positions and broadening observed by the authors at $\sim$1.5~\AA{} and $\sim$4~\AA{} (see \textbf{Figure~S6}). This agreement with literature data demonstrates that both models capture the local structural features of molten Li$_2$CO$_3$ reasonably well.

Upon increasing the temperature to 1300~K (\textbf{Figure~\ref{fig:rdfs}c}, the Li-C, C-O, C-C (magenta), and Li-O peaks begin to develop shoulders and further broaden, particularly beyond $\sim$3~\AA, $\sim$6~\AA, $\sim$5~\AA, and $\sim$4~\AA, respectively. This progressive broadening confirms that the melt structure becomes increasingly disordered with temperature. At 1500~K (\textbf{Figure~\ref{fig:rdfs}d} and \textbf{Figure~S4c}), the Li-O peak at $\sim$2~\AA, and the Li-C peak at $\sim$3~\AA{} show reduced intensity accompanied by pronounced shoulder formation. We do not observe any discernible peaks beyond $\sim$6~\AA, confirming the complete loss of long-range order and the molten state of Li$_2$CO$_3$. 

Compared to AIMD data at 1500~K (\textbf{Figure~S4c}), the MACE model exhibits similar RDFs (\textbf{Figure~\ref{fig:rdfs}d}) except a marginally more pronounced shoulder broadening for the Li-C peak at $\sim$3~\AA{} and a reduced broadening for the C-O and C-C peaks at $\sim$4~\AA{} and $\sim$5~\AA, respectively. On the other hand, the NequIP and MACE-MP-0 models correctly identify the peak positions in their predicted RDFs but fail to accurately reproduce the extent of peak broadening observed in AIMD across all temperatures (\textbf{Figures~S7-S9}). Nevertheless, both NequIP and MACE-MP-0 successfully capture the melting of Li$_2$CO$_3$ above 1000~K, as evidenced by the complete loss of structural ordering beyond $\sim$6~\AA{}.  

We calculated the total RDF ($g(r)$) for molten Li$_2$CO$_3$ at different temperatures to further examine the evolution of the Li$_2$CO$_3$ structure, as compiled in \textbf{Figure~S10}. In the range of $2-4$~\AA{}, the peaks exhibit a gradual shift and reduced peak intensity towards larger $r$ values with increasing temperature, indicating a loss of short range order in the structure. At distances beyond $\sim$6~\AA{}, the peaks become progressively broad and diminished at high temperatures (e.g., 1500~K), ultimately converging to $g(r)\approx 1$, which signifies the complete loss of long range order in the molten state.

\begin{figure}[h!]
    \centering
    \includegraphics[width=1\textwidth]{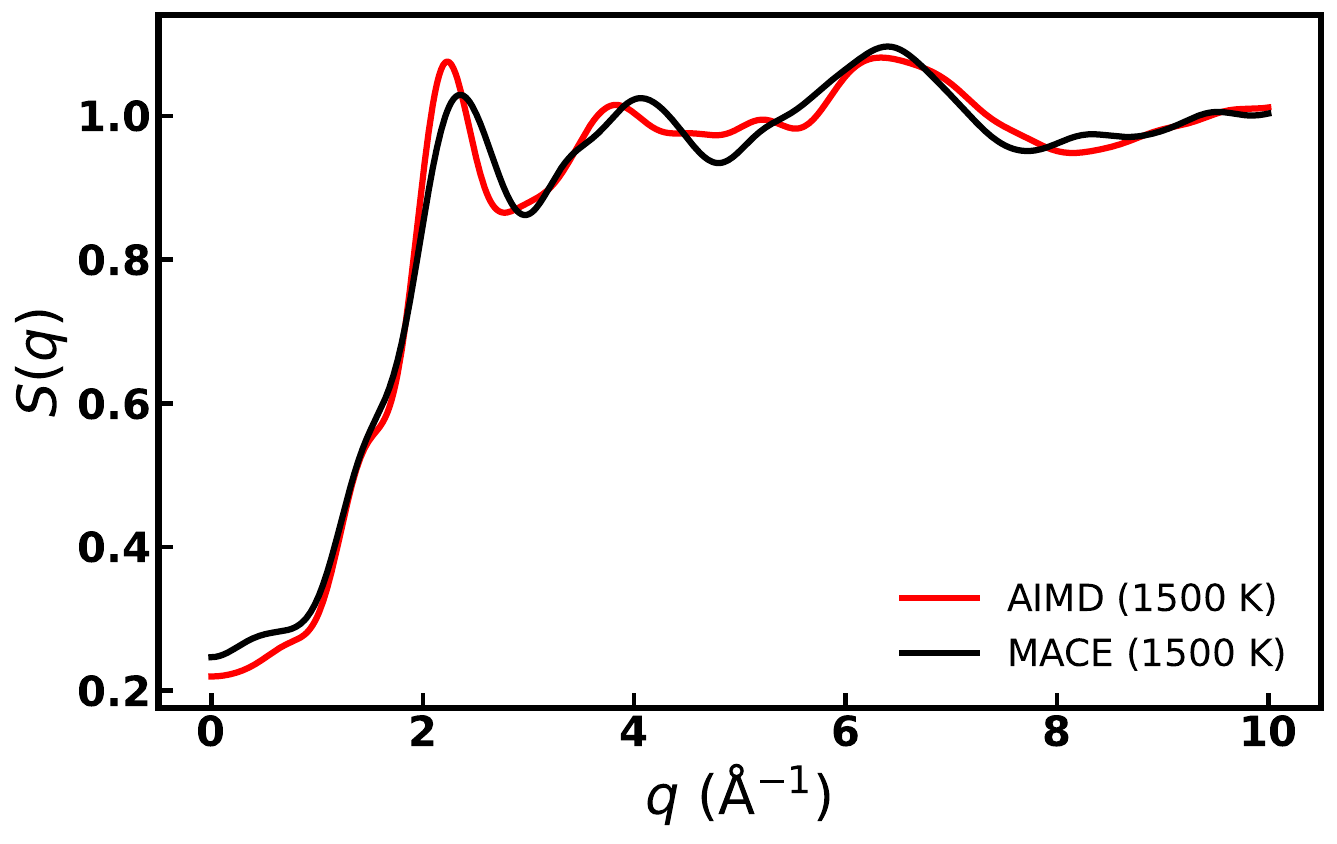}
    \caption{$S(q)$ for molten Li$_2$CO$_3$ at 1500~K, as calculated using AIMD (red) and MACE (black).}
    \label{fig:sq}
\end{figure}

As an additional validation of the structures generated by our MACE model, we calculate the $S(q)$ at 1500~K and compare that with the AIMD data in \textbf{Figure~\ref{fig:sq}}. The $S(q)$ predictions from the NequIP and MACE-MP-0 models at 1500~K are provided in \textbf{Figure~S11}. The detailed methodology used to compute $S(q)$ follows previous studies\cite{tovey2020dft,Wilding2016_LowDimNetwork_Na2CO3}, with the necessary parameters compiled in \textbf{Table~S3}. To further test the scalability and the accuracy of the MACE model, we benchmark the predictions by MACE at 1500~K in $2\times2\times2$ and $4\times4\times4$ supercells of the Li$_2$CO$_3$ structure (\textbf{Figure~S13}).

Importantly, the high degree of overlap between the AIMD (red) and the MACE (black) $S(q)$ curves in \textbf{Figure~{\ref{fig:sq}}} indicates that the MACE model successfully reproduces $S(q)$ and accurately captures the correct peak intensities as well. Additionally, we compare the MACE predictions for molten Li$_2$CO$_3$ at 1000~K with experimental $S(q)$ at 1013~K (see \textbf{Figure~S12})\cite{Sessa2024_LiKCO3_Polarization,Takeuchi2007_Structure_MoltenLi2CO3} and observe remarkable similarities between the MACE predicted and experimental $S(q)$ profiles, with minor deviations occurring beyond $\sim$5~\AA{}$^{-1}$. In the case of NequIP and MACE-MP-0, both models mostly capture the correct peak positions and peak broadening in $S(q)$ compared to AIMD, with deviations in the regions of $\sim$2.5~\AA{}$^{-1}$, 4.5~\AA{}$^{-1}$ and 6~\AA{}$^{-1}$, whereas MACE only shows deviations around 6~\AA{}$^{-1}$.

\subsection*{Transport properties in Li$_2$CO$_3$}
\begin{figure}[h!]
    \centering
    \includegraphics[width=1\textwidth]{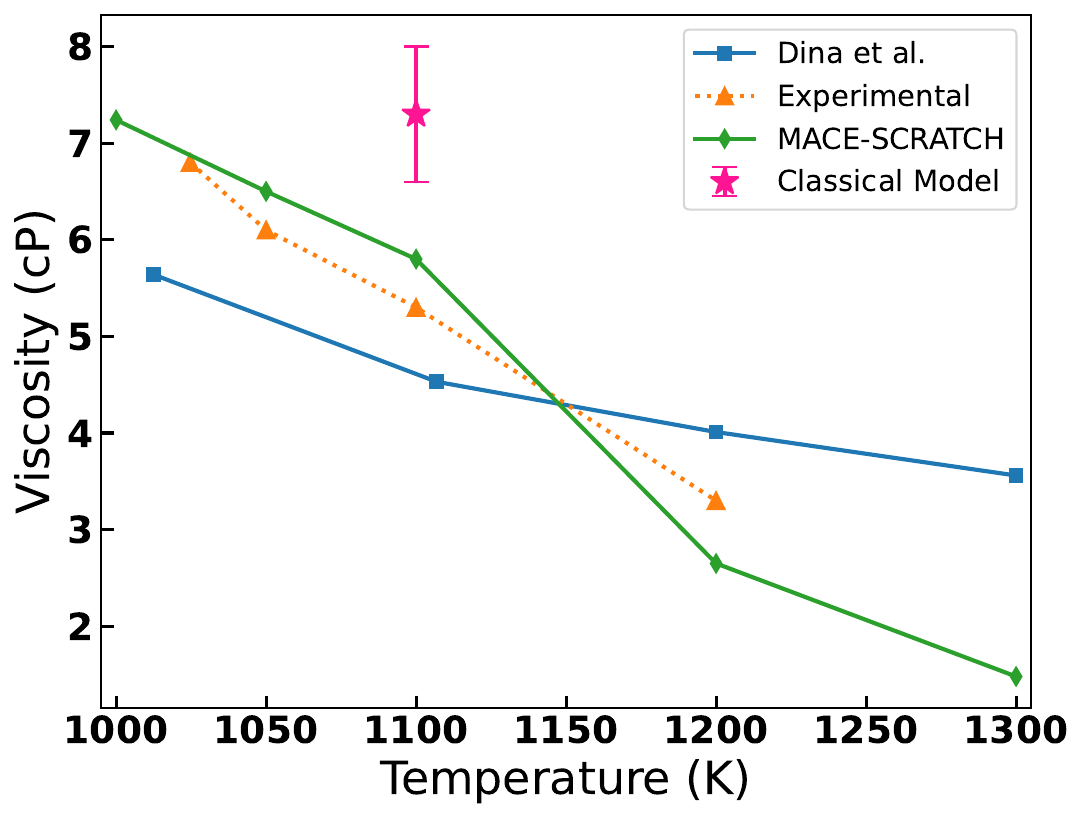}
    \caption{Viscosity of Li$_2$CO$_3$ (in units of centipoise or cP) as a function of temperature. Values predicted by the MACE-model (green) are compared with experimental data (orange), DeepMD model results (blue), and a classical force field (pink).}
    \label{fig:viscosity} 
\end{figure}

\noindent Given the crucial role that $\eta$ plays in the flow of any molten medium, we compare the MACE-computed $\eta$ at different temperatures with those from experimental\cite{Janz1988_MoltenSaltData} and  computational\cite{kussainova2023molecular,mondal2020genetic} studies in \textbf{Figure~{\ref{fig:viscosity}}}. We also summarize the MACE-computed $\eta$ in \textbf{Table~S4}. Importantly, the $\eta$ values predicted by our MACE model (green diamonds in \textbf{Figure~{\ref{fig:viscosity}}}) closely match the experimental trends (orange triangles\cite{Janz1988_MoltenSaltData}), and significantly outperform the DeepMD model (blue squares\cite{kussainova2023molecular}) in terms of reproducing experimental observations. Also, predictions from a classical parameterized force field\cite{mondal2020genetic} overestimates the experimental $\eta$ at 1100~K. Thus, our optimized MACE model is able to reproduce experimental trends in $\eta$ accurately compared to other computational models.

\begin{figure}[h!]
    \centering
    \includegraphics[width=1\textwidth]{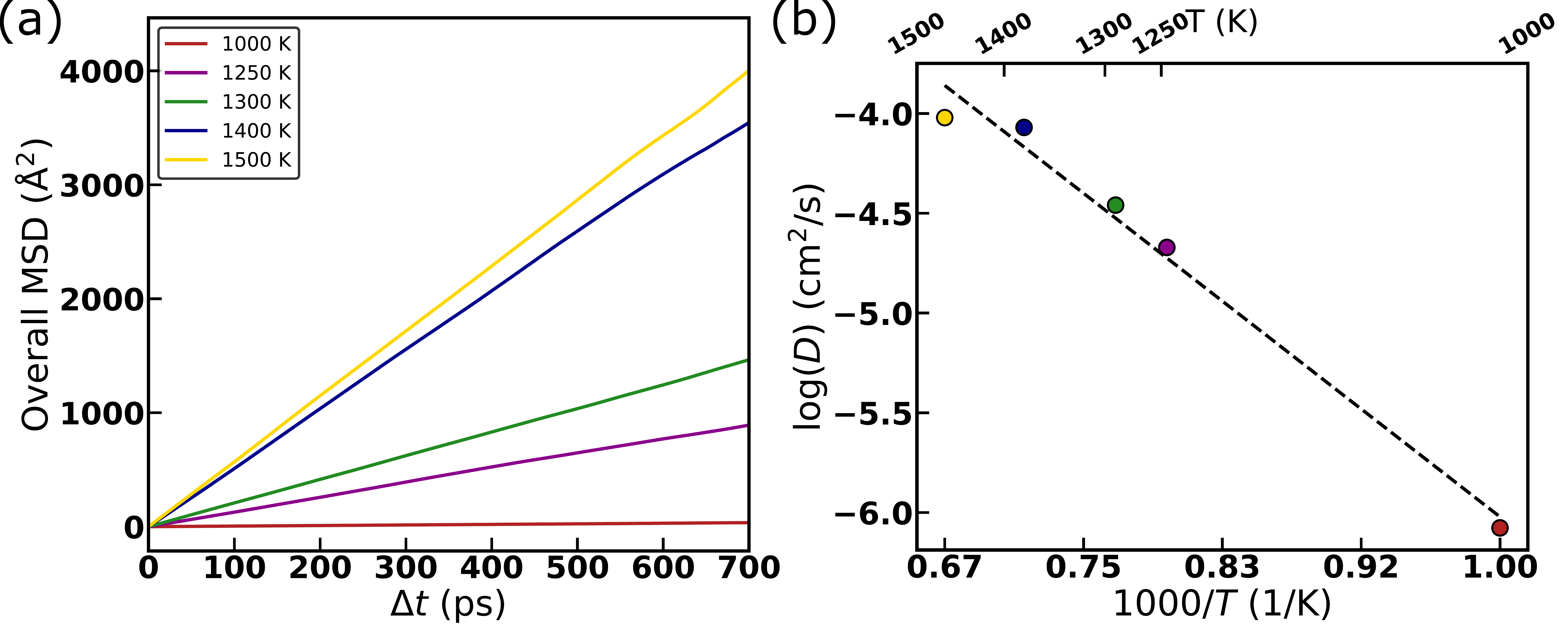}
    \caption{a) Mean-squared displacement (MSD) as a function of $\Delta t$, obtained from 700~ps MACE-MD simulations. Data are shown for five temperatures: 1500~K (yellow), 1400~K (blue), 1300~K (green), 1250~K (pink), and 1000~K (red). (b) Arrhenius representation of the diffusion coefficient, plotted as $\log D$ (with $D$ in~cm$^2$/s) versus $1000/T$. Symbols are color-coded according to panel~(a), and the dashed black line indicates the linear fit.}
    \label{fig:msd}
\end{figure}

We obtain the $D^*$ of Li$^+$ in Li$_2$CO$_3$ using long time-scale and large-system MD simulations, done with the MACE potential at various temperatures, and plot the results in \textbf{Figure~\ref{fig:msd}}. The overall averaged MSD of individual Li$^+$ combining the ionic displacements along all three Cartesian directions, as a function of $\Delta t$, is presented in \textbf{Figure~\ref{fig:msd}a}, with the yellow, blue, green, pink, and red curves indicating 1500~K, 1400~K, 1300~K, 1250~K, and 1000~K, respectively. \textbf{Figure~{\ref{fig:msd}}b} plots the $D^*$ (denoted simply as '$D$' in the figure) at individual temperatures, obtained from the MSD-$\Delta t$ slopes in panel a, with the color of the dots matching the colors used in panel a. The bottom and top x-axes of \textbf{Figure~{\ref{fig:msd}}b} correspond to 1000/$T$ and $T$, respectively, with the slope (dashed black line) proportional to the $E_a$. The raw MSD-$t$ curves containing the direction-dependent MSD of Li, at different temperatures, are compiled in \textbf{Figures S14} and \textbf{S15}.

At all temperatures, the MSD increases linearly with $\Delta t$, indicating diffusive behavior. Based on the MSD data, we calculate the Li$^+$ $D^*$ in molten Li$_2$CO$_3$ to be $8.38 \times 10^{-7}$~cm$^2$/s at 1000~K, $2.13 \times 10^{-5}$~cm$^2$/s at 1250~K, $3.47 \times 10^{-5}$~cm$^2$/s at 1300~K, $8.51 \times 10^{-5}$~cm$^2$/s at 1400~K, and $9.53 \times 10^{-5}$~cm$^2$/s at 1500~K, as plotted in \textbf{Figure~\ref{fig:msd}b}. The Arrhenius fit of $D^*$ with $T$ yields an activation energy of $E_{\mathrm{a}} = 1.287$~eV, with an $R^2$ score of 0.98. For comparison, earlier studies reported $E_{\mathrm{a}} \sim 1.34$~eV \cite{SEI6} in molten Li$_2$CO$_3$, while diffusion studies in bulk crystalline phases reported migration barriers in the range of $0.23-0.49$~eV \cite{iddir2010li}. The estimated statistical uncertainty, $D^* \pm \sigma$, as obtained via block-averaging of MSD-based diffusivities is summarized in \textbf{Table~S5}.

\textbf{Figure~S16} shows the temperature-dependent MSCD data, as a function of $\Delta t$ for molten Li$_2$CO$_3$ (panel~a) and the corresponding $H_R$ (panel~b), as extracted from MACE MD simulations on a $4\times4\times4$ supercell over 200~ps at 1000~K, 1300~K, and 1400~K. Importantly, $H_R$ remains well below unity even at high temperatures, with values of 0.40 at 1400~K, 0.31 at 1300~K, and 0.20 at 1000~K, indicating that Li-ion transport in molten Li$_2$CO$_3$ is fundamentally strongly correlated. Particularly, a $H_R$ of 0.20 (at 1000~K) indicates a significant departure from the ideal Nernst-Einstein behavior, signifying concerted migration as the dominant transport mechanism compared to uncorrelated site-to-site hopping of Li-ions. Note that the extent of correlated Li$^+$ motion in molten Li$_2$CO$_3$ is not immediately apparent from $D^*$ estimates of \textbf{Figure~{\ref{fig:msd}}a} and requires explicit calculations of MSCD and $H_R$. Within the ionic diffusion framework proposed by He \textit{et~al.}~\cite{he2017origin}, such concerted migration events typically arise when ions in high-energy sites hop downhill, thereby causing a possible reduction in the effective energy barrier for neighboring ions that move uphill.

Additionally, we find that molten Li$_2$CO$_3$ exhibits pronounced directional anisotropy in Li$^+~D^*$, as revealed by the direction-dependent MSD curves at lower temperatures (\textbf{Figures~S14} and \textbf{S15}). For example, at 1000~K, the $c$-axis of molten Li$_2$CO$_3$ supports faster Li transport ($D_c^\ast > D_b^\ast > D_a^\ast$, \textbf{Figure~S15}), whereas at 1400~K the three components converge to similar values (\textbf{Figure~S14b}), reflecting a transition towards isotropic diffusion at elevated temperatures. Our analysis of the transient oxygen-centered Voronoi cages (i.e., regions of space closer to oxygen than other ions) in molten Li$_2$CO$_3$, through which Li$^+$ migrate (\textbf{Figure~S17}), also supports the notion of anisotropic Li$^+$ motion at lower temperatures. For instance, at 1000~K (\textbf{Figure~S17a}), the $c$-axis displays markedly smaller fluctuations in the oxygen-centered Voronoi cage volumes, compared to the $a$ and $b$ axes, which indicates a more persistent local topology that preferentially facilitates Li$^+$ transport along the $c$-axis. On the other hand, at 1400~K (\textbf{Figure~S17b}), the cage volume fluctuations are similar along all three crystallographic directions consistent with isotropic Li$^+$ transport.

\begin{figure}[h!]
    \centering
    \includegraphics[width=1\textwidth]{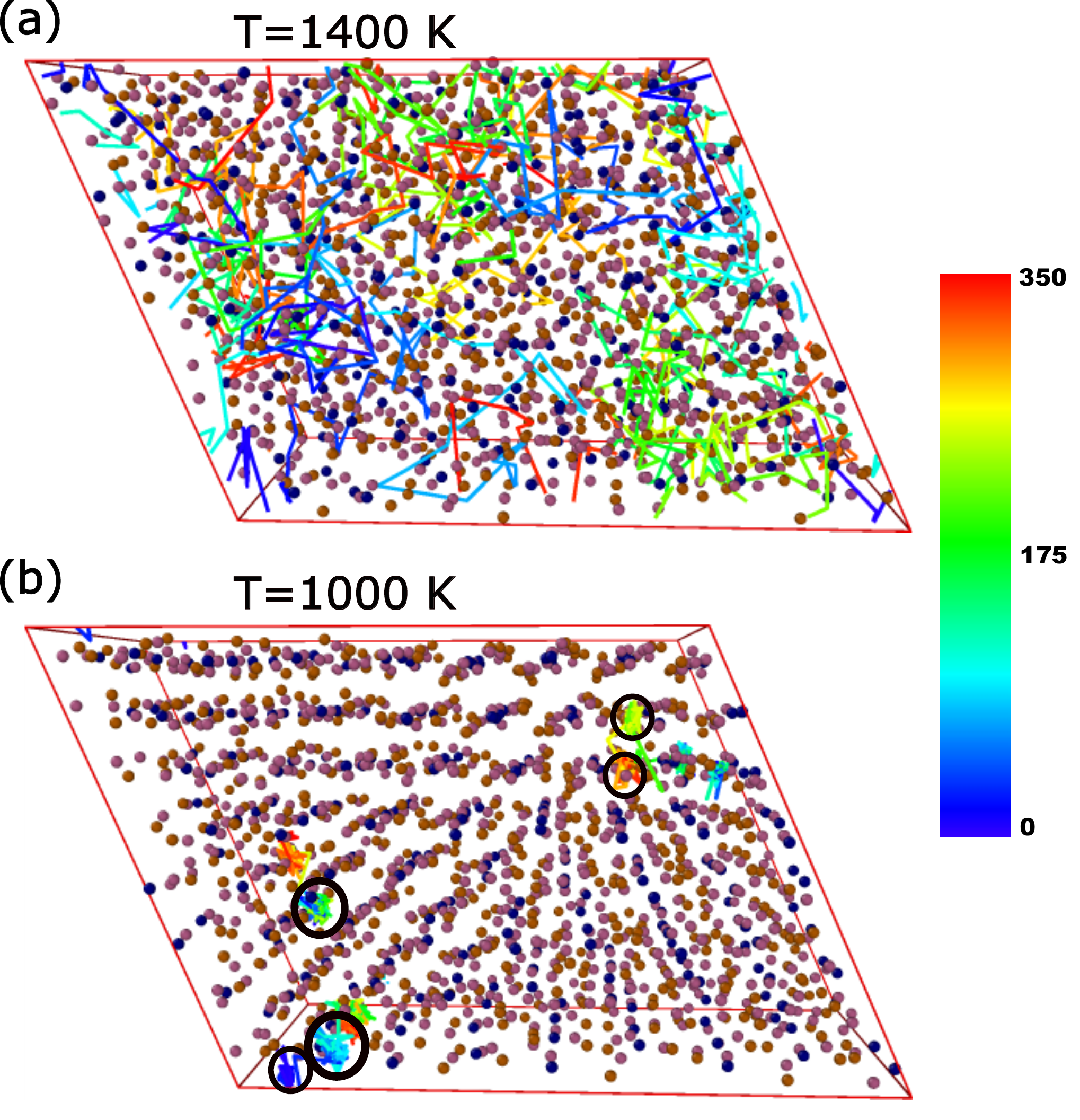} 
    \caption{(a) Unwrapped diffusion trajectories of representative Li$^+$ in molten Li$_2$CO$_3$ at 1400~K (panel a) and at 1000~K (panel b) over 350~ps. The rainbow color map traces an ion’s motion from blue (t = 0~ps) to red (t = 350~ps), depicting the temporal evolution of its path. The black circles at 1000~K mark regions of localized Li motion due to transient trapping by neighboring carbonate groups.}
    \label{fig:trajectories}
\end{figure}

To further understand the concerted nature of Li$^+$ motion, we visualize the unwrapped trajectories of representative Li$^+$ motion in molten Li$_2$CO$_3$ in \textbf{Figure~{\ref{fig:trajectories}}}, over 350~ps, at 1400~K (panel a) and at 1000~K (panel b). We unwrap the trajectory of motion across periodic images to improve visualization and interpretation of ionic motion. We employ a rainbow color map as a temporal marker, where blue and red denote the initial ($t$=0~ps) and final ($t$=350~ps) positions, respectively, of the ions. Importantly, the continuous nature of the Li$^+$ trajectories at 1400~K (\textbf{Figure~{\ref{fig:trajectories}}a}) reflects the extensive migration of ions through the molten carbonate environment and signifies lower correlation among Li$^+$ motion. In contrast, the black circles at 1000~K (\textbf{Figure~{\ref{fig:trajectories}}b}) highlight regions where Li$^+$ temporarily oscillate between nearby carbonate coordination sites (or cages), indicating transient trapping events before continuing its migration through the molten environment. Thus, the nature of the Li$^+$ concerted motion undergoes a significant change from 1000~K to 1400~K, which is also reflected in the increase in $H_R$ from 0.2 at 1000~K to 0.40 at 1400~K (\textbf{Figure~S16b}) and the increasing isotropy of Li$^+$ motion (\textbf{Figures~S14b} and \textbf{S15}) with increasing temperatures.

The MACE-calculated RDFs further support our observations on the transient trapping of Li$^+$ at lower temperatures. For example, we observe pronounced Li-C peaks at $\sim 3$~\AA{} and $\sim 5$~\AA{}, while the Li-O exhibits a strong peak at $\sim 3$~\AA{} at 1000~K (\textbf{Figure~{\ref{fig:rdfs}}b}). Notably, both the Li-C and Li-O peaks appear sharper at 1000~K than at 1400~K (\textbf{Figure~S5b}), indicating a more `ordered' local coordination environment and lower thermal disorder at 1000~K compared to 1400~K. Also, the sharper Li-C and Li-O peaks at 1000~K (than at 1400~K) suggest that Li$^+$ experience stronger coordination with nearby carbonate groups, reflecting the possibility of (transient) trapping of Li$^+$ by the neighboring carbonate groups. 

For elucidating the dynamic correlation among the Li-ions, we analyze the self ($G_s (r,t)$) and distinct ($G_d (r,t)$) parts of the van Hove correlation function at 1500~K, 1250~K, and 1000~K (\textbf{Figure~S18}). While $G_s(r,t)$ characterizes the displacement statistics of individual ions and is directly related to the MSD, $G_d(r,t)$ probes the relative motion and spatial correlations between distinct ions. At 1500~K, we observe bright horizontal bands in $G_d (r,t)$ that rapidly decay in intensity over a 2.5~ps window, consistent with liquid-like motion. In contrast, the horizontal bands at 1250~K and 1000~K remain sharper and persist for longer timescales, particularly in the 2-6~\AA{} range, indicating stronger transient order and more concerted motion, consistent with our previous observations in \textbf{Figures~{\ref{fig:trajectories}}}, \textbf{S14-S17}. We further calculate the ionic conductivity ($\kappa$), as obtained from the Nernst- Einstein relation, plot it as a function of temperature in \textbf{Figure~S19}, and compare the resultant activation energy with experimental data\cite{Janz1988_MoltenSaltData}.

\subsection*{Discussion}
\noindent In this work, we employed two equivariant graph-based MLIPs, NequIP and MACE, to understand the structure and Li-ion dynamics in molten Li$_2$CO$_3$, with broad applications in MCFCs and batteries. We trained both MLIPs on \textit{ab initio} melt-quench trajectories consisting of 3600 structures at 1500~K and 1988 structures at 1250~K (\textbf{Figure~{\ref{fig:workflow}}}). Notably, we found MACE to be the best-performing model, given its low errors on the test set (\textbf{Figures~S1} and \textbf{S2}), as quantified by energy and force errors of 0.2~meV/atom and 0.17~meV/\AA{}, respectively, on the 1000~K trajectory test set. Notably, our trained MACE model exhibited better performance than the foundational MACE-MP-0 model (without any fine-tuning) on our dataset. While fine-tuning MACE-MP-0 is a feasible strategy to accelerate MD simulations of specific systems, we trained the MACE architecture from scratch in our work to bias the model towards maximal accuracy towards the Li$_2$CO$_3$ system.

Remarkably, our best-performing MACE model requires only two message-passing layers, while the best-performing NequIP model requires five layers (\textbf{Figure~S21}), indicating the efficiency of the MACE model in learning the contributions of local bonding features to the overall potential energy surface.\cite{Batatia2023_MACE_JCP} In our benchmarking, we fixed the equivariant message-passing order to $L_{\mathrm{max}}=1$ for both models to ensure a consistent comparison. However, NequIP did not achieve our optimal MACE model's accuracy even when increasing the equivariant order to $L_{\mathrm{max}}=2$, indicating the superior featurization of the local geometry and the utility of including higher body-order interactions by the MACE architecture. 

Given that training both models typically requires GPU acceleration, we trained the models using a single NVIDIA V100 GPU. Subsequently, the optimized MACE model was able to simulate 760~ps of molten Li$_2$CO$_3$ containing 1536 atoms within 24 hours on a NVIDIA A100 GPU. For comparison, our previous study showed that the invariant MTP model, which is not based on graph networks, was easier to train and remained more computationally efficient for MD simulations on CPUs \cite{choyal2025exploration,choyal-sai}. Although MACE achieved higher accuracy than NequIP in our benchmarking, it was approximately five times slower to train than NequIP with a single GPU (\textbf{Figure~S21}). In CPU-based MD simulations of a 192-atom system using 16 threads on a single core, MACE was also significantly slower ($\approx 10\times$) than NequIP, highlighting the superior CPU performance of NequIP for production-scale MD (\textbf{Figure~S22}). Thus, our computational speed benchmarks on both CPUs and GPUs is further indicative of the speed-accuracy trade-off that users have to optimize for in training equivariant models, which also depends on the choice of hyperparameters and the specific system itself.

We characterized the structural properties of molten Li$_2$CO$_3$ using RDFs and $S(q)$ across multiple temperatures (\textbf{Figures~{\ref{fig:rdfs}}} and \textbf{\ref{fig:sq}}). AIMD simulations reveal a progressive loss of medium- and long-range structural order with increasing temperature, while the strong covalent C–O bonding motif remains intact throughout the melt. The MACE model reproduces both the RDFs and $S(q)$ with high fidelity, against both AIMD and experimental data (\textbf{Figures~S4}, \textbf{S11}, and \textbf{S12}) by accurately capturing the temperature dependence of peak positions and widths, and remains consistent upon scaling to larger system sizes. Furthermore, we computed the shear viscosity using MACE, which closely matches with available experimental measurements (\textbf{Figure~\ref{fig:viscosity}}), demonstrating that the model can faithfully recover critical properties of molten Li$_2$CO$_3$.

Analysis of Li$^+$ transport indicated that Li-ion diffusion in molten Li$_2$CO$_3$ proceeds via highly correlated motion, as evidenced by $H_R \ll 1$ over all temperatures (\textbf{Figure~{S16}}). While the extent of correlated motion is not immediately apparent from $D^*$ estimates (\textbf{Figure~{\ref{fig:msd}}}), the resultant $E_a$ are in agreement with previous simulations \cite{SEI6}. Notably, direction-dependent decomposition of Li$^+$ MSDs (\textbf{Figures~S14} and \textbf{S15}) reveal a pronounced transport anisotropy at 1000 K (with Li$^+$ diffusion preferred along the $c$-axis) and subsequent transition to a fully isotropic regime at 1400 K. Our visualization of the unwrapped Li$^+$ trajectories over 350~ps supports the notion of concerted motion and directional anisotropy, as indicated by transient trapping of Li$^+$ by carbonate cages at 1000~K (\textbf{Figure~{\ref{fig:trajectories}}}). The directional anisotropy of Li$^+$ motion can also be linked to fewer oxygen-centered Voronoi cage fluctuations along the $c$-axis at 1000~K (reflecting persistent topologies that favor directional Li$^+$ transport) compared to 1400~K (\textbf{Figure~S17}). Further, analysis of the van Hove correlation function points to a higher degree of dynamic correlation at 1000~K compared to liquid-like uncorrelated motion at 1500~K (\textbf{Figure~S18}). Thus, it is clear that the extent of correlation among Li$^+$ motion in molten Li$_2$CO$_3$ is highly temperature dependent and tends to reduce with increasing temperatures. Although longer simulations and additional temperature sampling would improve statistical accuracy, our calcuted MACE MD trajectories already exhibit clear temperature-dependent diffusion regimes, enabling reliable estimation of transport properties.

\subsection*{Conclusion}
\noindent In summary, we have performed a comprehensive atomistic investigation of the structure and transport properties of molten Li$_2$CO$_3$ using equivariant MLIPs, namely MACE and NequIP. Our initial benchmarking demonstrated that the MACE model provided superior accuracy and transferability compared to NequIP and the foundational MACE-MP-0 models, based on independent test set evaluations. The structural properties predicted by our optimized MACE model, such as the persistence of the C-O pair correlation even upon melting based on RDFs and $S(q)$, show remarkable agreement with experimental and \textit{ab initio} data. Also, our shear viscosity estimates in molten Li$_2$CO$_3$ are closer to experimental values than previous computational estimates. A central finding of our study is our prediction that Li$^+$ transport in molten Li$_2$CO$_3$ is fundamentally governed by highly correlated, concerted motion rather than simple, uncorrelated random hopping, as evidence by our $H_R$ estimates that point towards a substantial deviation from the ideal Nernst-Einstein behavior. Apart from the transient trapping of Li$^+$ by neighboring carbonate cages contributing to the correlated motion, we also observe a high degree of directional anisotropy in Li$^+$ transport at lower temperatures, as characterized by the fluctuations in oxygen-centered Voronoi cage volumes and computed van Hove correlation functions. Thus, our results represent a unique temperature-dependent transition in Li$^+$ transport in molten Li$_2$CO$_3$, from highly correlated and anisotropic diffusion at lower temperatures to less correlated and isotropic diffusion at higher temperatures. Our work also demonstrates the utility of equivariant graph-based MLIP architectures, such as MACE, which can effectively scale to systems containing thousands of atoms and reach nanosecond timescales while maintaining near-DFT accuracy, for modelling molten systems. We hope that the insights gained in this work are useful in the optimization of MCFCs and for understanding Li$^+$ transport in amorphous Li$_2$CO$_3$-based SEIs.

\subsection*{Acknowledgments}
\noindent G.S.G. acknowledges financial support from the Science and Engineering Research Board (SERB), of the Department of Science and Technology, Government of India, under sanction number IPA/2021/000007. D.D. acknowledges the Indian Institute of Science for academic support and the Shell fellowship for financial support. A.P. and A.N.K. acknowledge the support and resources provided by Shell International Exploration \& Production Inc. and Shell India Markets Pvt. Ltd.  A portion of the density functional theory calculations and molecular dynamics simulations showcased in this work were performed with the computational resources provided by the Supercomputer Education and Research Center, Indian Institute of Science. The authors thank Sharan Shetty, Neha Solanki, and Jason Williams at Shell for their support. 

\subsection*{Supporting Information}
\noindent Model hyperparameters, energy and force parity plots, energy fluctuations during equilibration, radial distribution functions, static structure factors, viscosity data, mean-squared-displacement and Li-ion diffusivity analysis, oxygen-centered Voronoi cage-volume analysis, van Hove correlation function calculations, ionic conductivity estimates, and computational efficiency benchmarks.

\subsection*{Data and code availability}
\noindent All computed data and models constructed in this work are available freely for all via our \href{https://github.com/sai-mat-group/molten-li2co3}{GitHub} repository.

\subsection*{Conflicts of interest}
\noindent The authors have no conflicts of interest to declare.
%\subsection*{Funding}

%\section*{Guidelines for References}

\newpage
\bibliographystyle{unsrt}
\bibliography{sample}

\end{document}